\documentclass{ws-procs9x6}

\newcommand{\be}{\begin{equation}}
\newcommand{\ee}{\end{equation}}
\newcommand{\beq}{\begin{eqnarray}}
\newcommand{\eeq}{\end{eqnarray}}
\newcommand{\noi}{\noindent}

\newcommand{\lsim}{\stackrel{<}{\sim}}
\newcommand{\upc}{\uppercase}
\newcommand{\bra}{\langle}
\newcommand{\ket}{\rangle}

\begin{document}

\title{Muonium Lifetime and Heavy Quark Decays\footnote{\upc{I}nvited talk present at the \upc{V}ernon \upc{H}ughes \upc{S}ymposium, \upc{Y}ale \upc{U}niversity, \upc{N}ovember 14--15, 2003.}\\ (Lessons Learned from Muonium)}

\author{William J. Marciano\footnote{\upc{W}ork supported by \upc{DOE G}rant \upc{DE-AC02-98CH10886}}}

\address{Brookhaven National Laboratory \\ Upton, New York\ \ 11973}

\maketitle

\abstracts{Environmental effects on the muon lifetime are described. A general theorem on the cancellation of bound state phase space suppression and final state interaction enhancement is illustrated for muonium and muonic atoms. Lessons from those bound muon examples are applied to the $b$ decay puzzle and apparent inconsistencies in $K_{e3}$ decay rates.}

\section{Introduction}

When asked to speak at this Symposium honoring the distinguished scientific career of Vernon Hughes, I knew the word ``muon'' would have to appear in my title and some property of that particle would provide my theme. Indeed, everyone at Yale knows the answer to I.I. Rabi's famous question, uttered when confirmation of the muon's existence was announced: ``Who ordered that?'' It must have been Rabi's student Vernon Hughes. Who else was able to forge such a distinguished research career based to a large measure on studies of or with that particle?

Better yet, why not talk about muonium. If Vernon Hughes liked anything more than the muon, it was muonium, the hydrogen like bound state of $\mu^+$ and $e^-$. He and his collaborators discovered muonium\cite{hughes} and lovingly studied its properties. Someday, I believe that muonium will have numerous as yet unimagined applications in physics, chemistry and life sciences,\cite{muon} a glorious tribute to its discovery. So, I decided to talk about the muonium lifetime. That may seem like a peculiar topic for an entire talk. Doesn't everyone know that muonium and the muon have the same lifetime? Well, they are almost the same, but not quite. In fact, why the difference is so small provides an instructive lesson on universal bound state principles that find applications in atomic, nuclear and particle physics; three communities that all claim Vernon Hughes as a favorite son. So, I think ``lessons learned from muonium'' is an appropriate theme  to share with this diverse audience that has gathered to recall memories of Vernon Hughes and his scientific legacy.

The contents of this talk are based on work done with Andrzej Czarnecki and G. Peter Lepage.\cite{czar} We studied modifications of the $\mu^+$ lifetime in matter due to muonium ($M=\mu^+e^-$) formation. Our study was motivated by an experiment underway at PSI\cite{carey} which is designed to measure the muon lifetime to one part per million. That experiment will use electronic timing technology developed for the muon $g-2$ measurement\cite{hughestwo} at Brookhaven; so, it can be considered a descendent of that heroic effort by Vernon Hughes. The question we addressed was how do bound state effects in muonium impact the muon lifetime? Is there a correction that must be made? To our surprise, the leading phase space suppression due to atomic binding turned out to be cancelled by final state $e^+e^-$ interactions that follow the decay process.  That cancellation was no accident and is not special to muonium decay. In fact, similar cancellations have been discussed in the literature for atomic screening of nuclear beta decay,\cite{serber} muonic atoms ($\mu^-N$),\cite{uberall} $b$ quark decays,\cite{bigi} etc. Here, I hope to describe the universal physics responsible for such cancellations. I will also discuss what small residual bound state effects do not cancel. They can be very accurately computed for muonium, making that bound state system a model for considerations of more complicated systems such as $b$ hadron lifetimes. So, my plan in this talk is to first discuss the utility of a precise muon lifetime measurement for testing the Standard Model and searching for ``new physics'' effects.\cite{marc} Then I describe our work on muonium decay.\cite{czar} Similar but more pronounced effects discovered by \"Uberall in 1960 for muonic atoms\cite{uberall} are then briefly explained.
An analogous situation exists for hadronic $b$ quark bound states. I try to explain, using muonium based physics, what analogous effect might be responsible for the unexpected short $\Lambda_b$ lifetime. Then I comment briefly on a little puzzle in the kaon system that is well known here at Yale\cite{kino} ``Why do charged and neutral $K_{e3}$ decay rates disagree with one another by about 9.5\% even after isospin violation corrections are applied?'' That situation has important implications for the determination of the CKM matrix element $|V_{us}|$ and tests of unitarity. Finally, I will end with some concluding remarks.

\section{The Muon Lifetime and Fermi Constant $G_\mu$}

Why measure the muon lifetime, $\tau_\mu$ with high precision? I recently discussed that issue in an article ``The Muon: A Laboratory for New Physics''\cite{marc} written in celebration of the 70th birthday of Alberto Sirlin (also a muon pioneer, but on the theory side). Here, I will briefly summarize some of the relevant points.

In the Standard Model, the only decay modes available to a free muon are $\mu^-\to e^-\bar\nu_e\nu_\mu$ and its radiative extensions $e^-\bar\nu_e\nu_\mu\gamma$, $e^-\bar\nu_e\nu_\mu\gamma\gamma$, $e^-\bar\nu_e\nu_\mu e^+e^-\dots$ The total inclusive decay rate, corresponding to the sum of those rates is related to the muon lifetime, $\tau_\mu$, and Fermi constant, $G_\mu$, via

\be
\Gamma(\mu^-\to e^-\bar\nu_e\nu_\mu(\gamma)) = \tau^{-1}_\mu = \frac{G^2_\mu m^5_\mu}{192\pi^3} f \left( \frac{m^2_e}{m^2_\mu}\right) (1+RC) \label{eqone}
\ee

\noi where

\be
f(x) = 1-8x + 8x^3 - x^4 - 12x^2\ell nx \label{eqtwo}
\ee

\noi is a phase space factor and $RC\equiv{}$radiative corrections includes virtual QED corrections to an effective four fermion interaction as well as bremsstrahlung and $e^+e^-$ emission. Currently, the RC are known to rather high order\cite{kino,vanrit}

\beq
RC & = & \frac{\alpha}{2\pi} \left(\frac{25}{4}-\pi^2\right) \left\{ 1+ \frac{\alpha}{\pi} \left( \frac23 \ell n \frac{m_\mu}{m_e}-3.7\right)  \right. \nonumber \\
& & \biggl.  + \left(\frac{\alpha}{\pi} \right)^2 \left(\frac49 \ell n^2 \frac{m_\mu}{m_e} -2.0 \ell n \frac{m_\mu}{m_e} + C \right) +\cdots \biggr\}
  \label{eqthree}
\eeq

\noi where $C$ is an unknown constant assumed to be $O (1)$. Comparing the expression in eq.~(\ref{eqone}) with the well measured muon lifetime\cite{hagi}

\be
\tau_\mu = 2.197035(40)\times10^{-6} sec \label{eqfour}
\ee

\noi gives the Fermi constant

\be 
G_\mu = 1.16637 (1)\times 10^{-5} {\rm ~GeV}^{-2} \label{eqfive}
\ee

\noi which is thereby known to 9ppm, making it the most precisely determined weak interaction parameter (not counting lepton masses). I note, that if more exotic decays exist (such as $\mu^-\to e^-\nu_e\bar\nu_\mu$, i.e.\ the wrong neutrinos), they should, in principle,  be subtracted out before $G_\mu$ is determined. Similarly, if the muon's environment allows additional decays, they should not be included (an example considered in section~4 is muon capture $\mu^- p\to \nu_\mu n$ in muonic atoms).

An experiment\cite{carey} under construction at PSI will aim for further improvement in $\tau_\mu$ and $G_\mu$ by a factor of 20! Future high intensity muon sources could probably push much further. However, on its own, $G_\mu$ is just a number. To utilize its precision requires a theoretical framework in which $G_\mu$ can be compared with other observables measured to comparable precision in a meaningful way. The Standard Model provides just such a renormalizable framework that relates $G_\mu$ to other electroweak parameters naturally. By natural, I mean that the radiative corrections are finite and calculable. Perhaps the best known example of such a relationship is\cite{sirlin}

\be 
G_F = \frac{\pi\alpha}{\sqrt{2} m^2_W (1-m^2_W/m^2_Z)(1-\Delta r)} \label{eqsix}
\ee

\noi where $\Delta r$ is an $O (\alpha)$ radiative correction that depends on $m_t$, $m_{Higgs}$ and, in principle any new physics that occurs in the quantum loop corrections to muon decay, $\alpha$, $m_Z$ and $m_W$. Notice that I refer to the Fermi constant as $G_F$ in eq.~(\ref{eqsix}). That is meant to suggest that $G_\mu=G_F$ in the Standard Model, but new physics would show up as a deviation, $G_\mu\ne G_F$, due to unaccounted loop effects in $\Delta r$.\cite{marc,marctwo} So, a strategy for testing the Standard Model is to determine $G_F$ with high precision in numerous ways, i.e.\ directly through $G_\mu$ and indirectly such as in eq.~(\ref{eqsix}), including known radiative corrections. That procedure currently restricts the Higgs mass $m_{Higgs}<180$ GeV and provides powerful constraints on dynamical symmetry breaking,\cite{marctwo} exotic muon decays,\cite{marc} properties of potential extra dimensions\cite{marctwo} etc. To illustrate the state of the art in those comparisons, I list in table~\ref{tabone} the current values of experimental quantities that are compared in relationships such as eq.~(\ref{eqsix}). Then I give in table \ref{tabtwo} the values of $G_F$ that follow from various inputs.\cite{marc}

\begin{table}[htb]
\tbl{Current experimental values for measured quantities that can be turned
 into Fermi constant determinations.}
{\begin{tabular}{@{}rcl@{}}
\hline
$\alpha^{-1}$&=&137.03599959(40) \\
$G_\mu$&=&1.16637(1)${}\times10^{-5}$ GeV$^{-2}$ \\
$m_Z$&=&91.1875(21) GeV \\
$m_W$&=&80.426(33) GeV \\
$\sin^2\theta_W(m_Z)_{\overline{MS}}$&=&0.23085(20) (Leptonic Asymmetries) \\
$\Gamma(Z\to\ell^+\ell^-)$&=&83.91(10) MeV \\
$\Gamma(Z\to\Sigma\nu\bar\nu)$&=&500.1(18) MeV \\
$\Gamma(\tau\to e\nu\bar\nu (\gamma))$&=&4.035(19)${}\times 10^{-13}$ GeV \\
$\Gamma(\tau\to\mu\nu\bar\nu(\gamma))$&=&3.933(19)${}\times10^{-13}$ GeV \\
\hline
\end{tabular}\label{tabone}}
\end{table}

\begin{table}[htb]
{\tbl{Various determinations of the Fermi constant using $m_t=174.3\pm5.1$ 
GeV and $m_{Higgs} =125^{+275}_{-35}$ GeV in the radiative corrections.\protect\cite{marctwo}}
{\begin{tabular}{@{}ll@{}}
\hline
$G_F$ (GeV$^{-2}$)${}\times10^5$ & Input + Rad. Corrections \\
\hline
1.16637(1) & $\tau_\mu$ \\
1.1700(70) & $\alpha, m_Z, m_W$ \\
1.1661(20) & $\alpha, m_W, \sin^2\theta_W(m_Z)_{\overline{MS}}$ \\
1.1672(20) & $\alpha, m_Z, \sin^2\theta_W(m_Z)_{\overline{MS}}$ \\
1.1650(20) & $\Gamma(Z\to\ell^+\ell^-), m_Z, 
\sin^2\theta_W(m_Z)_{\overline{MS}}$ \\
1.1666(45) & $\Gamma(Z\to\Sigma\nu\bar\nu), m_Z, 
\sin^2\theta_W(m_Z)_{\overline{MS}}$ \\
1.1666(28) & $\Gamma(\tau\to e\nu\bar\nu (\gamma))$ \\
1.1679(28) & $\Gamma(\tau\to\mu\nu\bar\nu(\gamma))$ \\
\hline
\end{tabular}\label{tabtwo}}}
\end{table}

Table \ref{tabone} illustrates the fact that $\alpha, G_\mu$ and $m_Z$ are all measured with exteme precision. However, the other quantities that they must be compared with to gleam information about $m_{Higgs}$ or new physics are not known nearly as well. That point is also clear in table~\ref{tabtwo}, where the current 9ppm value of $G_\mu$ is contrasted with the other determinations of the Fermi constant which have not yet reached $\pm0.1\%$ accuracy. A good goal for the future would be to push the other $G_F$ determinations to $\pm0.01\%$. At that level, they constrain the Higgs mass to about $\pm5\%$, could see supersymmetry loop effects, might uncover effects due to extra dimensions etc.

But why measure $G_\mu$ to better than $\pm0.0001\%$ via the muon lifetime? It is already 100 times better known than the $G_F$ with which it is compared. One answer is: It is a fundamental quantity and should be, therefore, measured as precisely as possible. A second answer might be: Why not stay a few steps ahead of the other measurements. However, there is a more persuasive argument based on current physics requirements, in particular,  the study of $\mu^-$ capture in hydrogen.

The basic point is that comparing the $\mu^+$ and $\mu^-$ lifetimes in matter provides indirect determinations of the $\mu^- N\to \nu_\mu N^\prime$ capture rates via the relation

\be
\Lambda_{capture} = \tau^{-1}_{\mu^-} - \tau^{-1}_{\mu^+} \label{eqseven}
\ee

\noi That has been the traditional way of obtaining capture rates for various elements.\cite{muon} It works very well for high $Z$ nuclei where the capture rate is large (it grows roughly as $Z^3\sim Z^4$), but becomes very difficult for hydrogen where one expects

\be
\Lambda_c (p\mu^-\to n\nu_\mu) \simeq 10^{-3} \Gamma(\mu\to e\nu\bar\nu) \label{eqeight}
\ee

\noi Of course, capture on hydrogen is the simplest and most interesting capture process. Furthermore, a puzzle exists from past studies of muon capture in hydrogen. The induced pseudoscalar constant $g_p$, has been directly measured in radiative muon capture\cite{gorr} and found to be about

\be 
g^{exp}_p \simeq 12; \label{eqnine}
\ee

\noi however, chiral perturbation theory predicts

\be
g^{th.}_p = 8.1\pm0.3 \label{eqten}
\ee

\noi The discrepency can be resolved by measuring $\Lambda_c$ for hydrogen using the $\mu^-$-$\mu^+$ reciprocal lifetime difference and extracting $g_p$ from the total capture rate. Since it only contributes a fraction of the capture rate, one must measure $\Delta\Lambda_c /\Lambda_c$ to $\pm1\%$ to determine $g_p$ to $\pm7\%$. That will require a 10ppm measurement of $\tau_{\mu^-}$ in hydrogen (now an approved PSI experiment\cite{adam,bardin}) along with the 1ppm $\tau_{\mu^+}$ experiment mentioned before.\cite{carey} So, in my opinion, the $g_p$ puzzle currently provides the main near term motivation for high precision muon lifetime measurements.

The PSI experiments will stop both the $\mu^+$ and $\mu^-$ and then basically count the decays as a function of time. The $\mu^-$ stopping material (hydrogen) will reduce the lifetime due to the capture mode, but what about modifications of the ordinary decay mode $\mu^-\to e^-\bar\nu_e\nu_\mu$ due to nuclear binding effects. Similarly, will bound state effects due to muonium $M=\mu^+ e^-$ formation in the stopping material affect the $\mu^+$ lifetime at 1ppm? These issues have been addressed in the literature.\cite{czar,uberall,gilin} In the next two sections I summarize what are rather surprising and interesting results from those studies.

\section{Muonium $(\mu^+e^-$) Decay}

Measurement of the $\mu^+$ lifetime (at rest) requires stopping the $\mu^+$ in matter and counting the number of outgoing muons as a function of time. At the 1ppm, one might worry about environmental effects on the lifetime. For example, will electron screening or muonium $(\mu^+e^-)$ formation modify the decay rate? The easiest case to examine is muonium, a simple bound state with very well defined properties.\cite{hughesthree}

Modifications of $\tau_\mu$ for the $1S$ bound state of muonium, due to Coulombic interactions, can be expressed as an expansion in terms of the two small dimensionless parameters $\alpha=1/137$ and $m_e/m_\mu\simeq 1/207$. Such effects must vanish as $\alpha\to0$ or $m_e/m_\mu\to0$. Hence, one expects corrections of the form $\alpha^n(m_e/m_\mu)^m$ where $n$ and $m$ are integers. Before considering the leading corrections, let me review some basic properties of muonium's ground state. The binding energy, average potential and kinetic energy, as well as average electron and muon velocities are given by

\beq
\qquad\qquad\qquad\qquad\qquad & E=\frac12 \alpha^2 m_e \simeq -13.5 {\rm~eV} \qquad\quad\qquad\qquad\qquad &(11a) \nonumber \\
&\langle V\rangle = -\alpha^2 m_e\simeq -27 {\rm~eV}\qquad\quad\qquad\qquad\qquad &(11b) \nonumber \\
&\langle T\rangle = \frac12 \alpha^2 m_e\simeq 13.5 {\rm~eV} \qquad\quad\qquad\qquad\qquad&(11c) \nonumber \\
&\langle \beta_e\rangle \simeq \alpha \quad\qquad\qquad\qquad\qquad\quad\qquad\qquad\qquad &(11d) \nonumber \\
&\langle \beta_\mu\rangle \simeq \alpha\frac{m_e}{m_\mu}  \quad\qquad\qquad\qquad\quad\qquad\qquad\qquad &(11e) \nonumber
\eeq

\addtocounter{equation}{1}

\noi The muon rest frame and lab frame differ because $\bra \beta_\mu\ket \ne0$. That effect gives some spectral distortion due to Doppler smearing of the positron energy by terms of $O(\alpha m_e/m_\mu)$. However, Lorentz invariance tells us that there are no corrections to the lifetime linear in velocity. Instead, that small muon lifetime velocity only gives rise to a very small time dilation increase

\be
\tau_M =\tau_\mu \left(1+\frac12 \bra\beta_\mu\ket^2\right) \simeq \tau_\mu \left(1+\frac12 \alpha^2m^2_e/m^2_\mu\right) \label{eqtwelve}
\ee

\noi which is about $6\times10^{-10}$, a negligible increase.

The next-to-leading spectral distortion will entail effects of order $-\bra V\ket/m_\mu = \alpha^2 m_e/m_\mu$. Naively, one might expect a potential energy shift in the outgoing positron energy to give rise to a lifetime reduction due to the phase space change

\be
m^5_\mu \to (m_\mu +\bra V\ket)^5 \to m^5_\mu (1-5\alpha^2 m_e/m_\mu) \label{eqthirteen} 
\ee

\noi That small $\simeq 1\times10^{-6}$ reduction would, if real, be about the same size as the expected PSI experiment's sensitivity.\cite{carey} However, it has been shown\cite{czar} that final state $e^+e^-$ interactions give rise to an equal but opposite sign effect that cancels the shift in eq.~(\ref{eqthirteen}). In fact, electromagnetic gauge invariance, can be used to prove a general theorem:\cite{czar} There are no $\alpha^n m_e/m_\mu$ corrections! The absence of such $1/m_\mu$ corrections is very well known to people who work on $b$ quark physics where the operator product expansion is used to show that there are no $1/m_b$ corrections\cite{bigi,ural} to the $b$ quark lifetime when placed in different hadronic bound state environments, i.e.\ $B_d$, $B_u$, $\Lambda_b$ etc. In other words, through first order in $1/m_b$ all $b$ hadrons should have the same lifetime, a somewhat anti-intuitive feature. I return to this point in section~5. Note also that the cancellation of phase space effects due to the electron screening potential with final state interactions also occurs for nuclear beta decays.\cite{rose} but is usually disguised because the standard $\beta$-decay formalism employs $Q$ values for atoms rather than fully ionized nuclei. That approach hides the phase space effect; so only the final state interaction is corrected for. It has also been observed for the decay of the $\mu^-$ while in a muonic  atom bound state. Indeed, the cancellation is a universal phenomenon.

A simple illustration of the above cancellation is provided by a static sphere model in which one thinks of muonium as a $\mu^+$ surrounded by a fixed thin electron sphere with radius${}\sim{}$Bohr radius. The effect of that sphere gives rise to a potential $V=-\alpha^2 m_e$ which shifts the entire positron decay spectrum by $V$. The fully integrated decay rate $\Gamma(\mu^+\to e^+\nu\bar\nu)$ for a $\mu^+$ in the sphere is modified such that

\be
\Gamma(\mu^+\to e^+\nu\bar\nu) = \int^{\frac{m^2_\mu+m^2_e}{2m_\mu}+V}_{m_e+V} dE \frac{d\Gamma(E-V)}{dE} \label{eqfourteen}
\ee

\noi There is an apparent change in phase space (limits of integration) which is cancelled by the final state interaction of the $e^+$ with the electron sphere which shifts the differential decay rate. Overall, the shifts in eq.~(\ref{eqfourteen}) correspond to a simple change of variables with no net effect.

So, we have learned that the very small time dilation shift in eq.~(\ref{eqtwelve}) is in fact the leading correction and it represents a totally negligible effect for 1ppm muon lifetime measurements. A similar argument can be made for electron screening effects in metals which collectively give rise to potentials similar to muonium.

What are some of the higher order (in $\alpha$ and $m_e/m_\mu$) corrections to the muonium lifetime? That question is not of phenomenological importance for the muon lifetime issue, but muonium is such a simple system that we should be able to learn some useful lessons from such a study. One such correction is due to the muon-electron hyperfine interaction which modifies the lifetime by terms of order $\alpha^4 m^2_e/m^2_\mu$. The QCD analog of that effect provides a leading $1/m^2_b$ correction to $b$ hadron lifetimes (along with time dilation), see section 5.

What are the leading $1/m^3_\mu$ corrections? Muonium exhibits two such effects. The first is the availability of a capture mode $M\to\nu_e\bar\nu_\mu$ in muonium\cite{czar} 

\be
\Gamma(M\to\nu_e\bar\nu_\mu) = 48\pi \left(\frac{\alpha m_e}{m_\mu}\right)^3 \Gamma(\mu\to e\nu\bar\nu) \label{eqfifteen}
\ee

Although that effect is tiny $\sim6.6\times10^{-12}$, it demonstrates an important feature. The 2 vs 3 body final state gives rise to a very large $48\pi$ enhancement fact. This type of capture effect can be quite important for $b$-hadron lifetimes where 2 body annihilation or scattering can play a significant role. A final order $(\alpha m_e/m_\mu)^3$ muonium decay correction is well illustrated by the simple example in eq.~(\ref{eqfourteen}). Classically, the entire range of integration down to $m_e$--$\alpha^2 m_e$ is allowed. However, the real lowest energy state positronium has only $\frac12 V$. So, part of the lower range of the integration in eq~(\ref{eqfourteen}) are actually not allowed. That effect is the real quantum phase space reduction. It suppresses the muon decay rate in muonium by a tiny correction $\sim-16(\alpha m_e/m_\mu)^3\simeq -7\times10^{-13}$. Although very small, it also exhibits a large factor $\sim16$ enhancement factor.

Overall, one finds the following lessons learned from the relationship between the muon and muonium total decay rates\cite{czar}

\beq
\Gamma(M\to{\rm all}) & = & \Gamma(\mu^+\to e\bar\nu_e\nu_\mu) \left\{ 1-\frac12 \biggl(\alpha\frac{m_e}{m_\mu}\right)^2 + O (\alpha^4 m^2_e/m^2_\mu) \biggr. \nonumber \\
& & \biggl. +48\pi (\alpha m_e/m_\mu)^3 - 16(\alpha m_e/m_\mu)^3 +\cdots \biggr\} \label{eqsixteen}
\eeq

\noi where the 4 corrections exhibited correspond to time dilation, hyperfine effects, annihilation and quantum phase space reduction respectively. Similar types of corrections will occur for other bound states, as we now illustrate by several examples.

\section{$\mu^-$ Decay in Atomic Orbit}

Muonic atoms have much larger Coulomb interactions than muonium because of their small radii and tight binding to nuclei. The average bound state potential and muon velocity are 

\beq
\bra V\ket & = & -(Z\alpha)^2 m_\mu \nonumber \\
\bra \beta_\mu\ket & = & Z\alpha \label{eqseventeen}
\eeq

For high $Z$ nuclei, $\beta_\mu$ will give rise to large spectral distortions. Again, the naive phase space reduction due to $\bra V\ket$ cancels with the electron-nucleus final state interaction, as observed by \"Uberall in 1960.\cite{uberall} The leading remaining effect is a time dilation increase of the $\mu^-$ in orbit lifetime by a $1+\frac12 Z^2\alpha^2$ factor which can be important for high $Z$ as well as a 10ppm $\mu^-$ lifetime measurement in $Z=1$ hydrogen. There are also capture processes\cite{bardin} which provide a new decay mode which grows significantly with $Z$,

\be \Lambda_{\rm capture}\simeq 1000 (Z\alpha)^3 \Gamma(\mu\to e\nu\bar\nu) \label{eqeighteen}
\ee

\noi and again has the large 2 body enhancement factor. (It actually grows as $Z^4$ modulo final state Pauli exclusion.) There is also a quantum state reduction of $O(Z^3\alpha^3)$ due to the suppression\cite{gilin} of decays with $E_{e^-} < Z^2\alpha^2 m_\mu$. 

Another interesting effect is decays with $E_{e^-}> \frac{m^2_\mu+m^2_e}{2m_\mu}$ due to the possibility of nuclear recoil. In fact, the $e^-$ spectrum will have a tail extending all the way to $m_\mu(1-Z^2\alpha^2)$. All such corrections are calculable;\cite{gilin} so they should not cloud the interpretation of the $\mu^-$ lifetime in hydrogen. Nevertheless, for a 10ppm experiment they must be closely scrutinized.

\section{$b$-Hadron Lifetimes}

A very nice illustration of the lessons learned from muonium is provided by the lifetimes of $b$ mesons and baryons. QCD rather than QED bound state potentials are involved, but the physics is universal. To a good first approximation all $b$-hadrons should have the same ($b$ quark) lifetime. That is a remarkable feature when one considers the broad range of $b$-hadron masses and the complexity of final state decay interactions. Nevertheless, the cancellation of naive phase space effects and QCD final state interactions will occur due to QCD gauge  invariance. As a result, there are no $1/m_b$ corrections to the $b$ quark lifetime induced by its hadronic environment. That property is well known to $b$ physics workers. It is usually proven by using the operator product expansion\cite{ural,volo} to show the leading lifetime corrections are of order $1/m^2_b$.

There are $O(1/m^2_b)$ time dilation corrections (due to the $b$ quark Fermi motion) as well as QCD hyperfine interactions that vary from one $b$ hadron to another. However, those corrections are expected to represent only few percent effects. Of course, the $1/m^3_b$ corrections to lifetimes can be potentially important. Two body processes $b+u\to c+d$ in $b$ baryons and $b+\bar d \to c+\bar u$ in the $B_d$ meson are relatively suppressed by $1/m^3_b$ but can have large 2 body phase-space enhancements similar to the $48\pi$ factor observed for $M\to\nu_e\bar\nu_\mu$ in section~3. But, are those corrections enough to bring all $b$-hadron lifetimes into accord with one another? To illustrate the current situation, I give in table~\ref{tabthree} some $b$ hadron properties along with their lifetimes\cite{hagi}

\begin{table}[htb]
\tbl{$b$-hadron properties and lifetime ratios.}
{\begin{tabular}{@{}cccc@{}}
\hline
State & Mass (MeV) & Lifetime (ps) & Lifetime/$\tau_{B^0_d}$ \\
\hline
$B^0_d = b\bar d$ & 5279 & 1.542(16) & 1 \\
$B^-_u=b\bar u$ & 5279 & 1.674(18) & 1.083(17) \\
$B_s = b\bar s$ & 5370 & 1.461(57) & 0.947(38) \\
$\Lambda_b=bud$ & 5624 & 1.229(80) & 0.797(53) \\
\hline
\end{tabular}\label{tabthree}}
\end{table}

\noi How do those lifetime ratios compare with theoretical expectations, after time dilation, hyperfine interactions, and 2 body ``capture'' interactions are taken into account?\cite{ural,volo} The predicted ratios are\cite{hagi}

\beq
\quad\qquad\qquad\qquad & \tau_{B^-}/\tau_{B^0_d} = 1+0.05 (f_B/200{\rm ~MeV})^2 \quad\qquad\qquad\qquad &(19a) \nonumber \\
\quad\qquad\qquad\qquad & \tau_{B_s}/\tau_{B^0_d} = 1\pm0.01 \qquad\qquad\qquad\qquad\qquad\qquad\qquad &(19b) \nonumber \\
\quad\qquad\qquad\qquad & \tau_{\Lambda_b}/\tau_{B^0_d} =0.9 \qquad\qquad\qquad\qquad\qquad\qquad\qquad\qquad &(19c) \nonumber
\eeq

\addtocounter{equation}{1}

\noi Those theoretical expectations seem to be in some disagreement with the lifetimes observed for $B_s$ and $\Lambda_b$. In fact, the direction of the disagreement appears to be correlated with their larger masses, i.e. looks like a phase space effect. The difference between theory and experiment, particularly for the $\Lambda_b$ is sometimes called the $b$ lifetime puzzle. How will it be resolved? Perhaps $\tau^{exp}_{\Lambda_b}$ and $\tau^{exp}_{B_s}$ will change. Maybe the $O(1/m^3_b)$ 2 body rates are larger than theory estimates. Large $O(1/m^4_b)$ effects may be the source.\cite{gabb} Here, I would like to suggest that a lesson learned from muonium may be the cause. The quantum phase-space reduction may be giving additional different $1/m^3_b$ corrections for the various $b$-hadrons. (Sometimes called a breakdown of quark-hadron duality by $b$ physics workers.) It is a tiny effect for muonium but could be a few percent for the $b$-hadron system. How this $b$ lifetime puzzle issue is resolved will be interesting to watch.

\section{The $K^+\to\pi^0 e^+\nu_e$ vs $K^0\to\pi^- e^+\nu_e$ Decay Puzzle}

Another interesting puzzle has recently surfaced in $K_{e3} (K\to\pi e\nu)$ decays. That special decay is important because it is traditionally used to obtain the CKM quark mixing matrix element $|V_{us}|=\sin\theta_{Cabibbo}$. The neutral $K_L\to\pi^+ e^-\bar\nu_e + K_L\to\pi^-e^+\nu_e$ decay rates together give rise to\cite{hagi}

\be
\Gamma(K_L\to\pi e\nu) = 0.494(5)\times10^{-14} {\rm ~MeV} \label{eqtwenty}
\ee

\noi Those relatively old results seem to have been recently confirmed by the KLOE collaboration\cite{fran} at the $\phi$ factory in Frascati; so, they should be taken seriously. On the other hand, a recent measurement by the E865 collaboration\cite{sher} at Brookhaven found

\be
\Gamma(K^+\to\pi^0 e^+\nu_e) = 0.273(5) \times10^{-14} {\rm ~MeV} \label{eqtwentyone}
\ee

\noi If isospin were a perfect symmetry, one would expect the ratio of those two rates to be 2 (because 2 modes are included in eq.~(\ref{eqtwenty})). However, one currently finds

\be
\frac{\Gamma(K_L\to\pi^\pm e\nu)}{\Gamma(K^+\to\pi^0 e\nu)} = 1.81(2)(3) \label{eqtwentytwo}
\ee

\noi a significant 9.5\% deviation from the isospin limit. That difference gives rise (even after isospin violating corrections) to quite different determinations\cite{marcthree} of $|V_{us}|$ from neutral and charged $K_{e3}$ decays, an unacceptable situation. 

What does the above problem have to do with lessons from muonium? Well, there are 2 sources of isospin violation in eq.~(\ref{eqtwentytwo}), the $m_d-m_u$ mass difference and QED (electromagnetic) effects. The first of those gives rise to a factor of\cite{leut}

\be
1+\frac32 \frac{m_d-m_u}{m_s-(m_d+m_u)/2} \simeq 1.034 \label{eqtwentythree}
\ee

\noi correction (from $\pi$-$\eta$ mixing) that is accounted for in extractions of $|V_{us}|$ from $K^+_{e3}$. It is not enough to account for the 9.5\% difference on its own. In fact, it is essentially cancelled by the $m_{K_L}-m_{K^+}$ mass difference effect which is primarily (but not totally) due to $m_d-m_u$. That gives rise to a 

\be
(m_{K^+}/m_{K_L})^5 \simeq 0.96 \label{eqnewtwentyfour}
\ee

\noi  compensating correction. The 9.5\% difference must be due to electromagnetic effects or experimental error(s).

Now is where the lesson from muonium comes in; specifically, the cancellation between electromagnetic phase space effects and final state interactions. Of the kaon and pion mass differences $m_{K_L}-m_{K^+}$, $m_{\pi^+}-m_{\pi^0}$\cite{hagi} involved in the $K_{e3}$ decays

\beq
m_{K^+} & = & 493.65 {\rm ~MeV} \nonumber \\
m_{\pi^0} & = & 134.97 {\rm ~MeV} \nonumber \\
m_{K_L} & = & 497.67 {\rm ~MEV} \label{eqtwentyfour} \\
m_{\pi^\pm} & = & 139.57 {\rm ~MeV} \nonumber 
\eeq

\noi the latter is  primarily of electromagnetic origin. It gives rise to a phase-space reduction of the expected ratio of 2 in eq.~(\ref{eqtwentytwo}) by\cite{leut}

\be
{\rm Pion~Mass~Phase~Space~reduction} \simeq \frac{0.1561}{0.1605} = 0.9726 \label{eqtwentyfive}
\ee

\noi That 2.6\% change goes in the right direction. However, there is a fairly large final state electromagnetic (Coulombic) interaction (FSI) between the $\pi^+ e^-$ (or $\pi^- e^+$) which enhances the neutral $K_{e3}$ relative to the charged $K_{e3}$ by 

\be
{\rm FSI} \simeq 1 +\pi\alpha \simeq 1.023 \label{eqtwentysix}
\ee

\noi Multiplying eq.~(\ref{eqtwentysix}) times eq.~(\ref{eqtwentyfive}) demonstrates the near cancellation as expected from the general theorem found for muonium, beta-decay, $b$ decays etc. It is again a consequence of electromagnetic gauge invariance ; albeit a more subtle demonstration. In a sense, the cancellation confirms the claim made above that the pion mass difference is primarily of electromagnetic origin. Where does that leave us? The product of the 4 isospin violating corrections give rise to an overall factor of 

\be
(1.034\cdot 0.96)/(0.9726\cdot 1.023) = 0.998 \label{eqtwentyseven}
\ee

\noi The 9.5\% deficit remains and leads to about a 4.7\% difference in the values of $|V_{us}|$ extracted from charged and neutral $K_{e3}$ results. Something is wrong and it doesn't appear to be the up-down mass difference or the electromagnetic effects (since the theorem works). There are various possibilities or some combination of them: 1) One of the $K_{e3}$ results is incorrect. It could be an actual branching ratio measurement or the use of an incorrect $K^\pm$ or $K_L$ lifetime employed to convert to the partial decay rates in eqs.~(\ref{eqtwenty}) or (\ref{eqtwentyone}). In the case of both $K^\pm$ and $K_L$, their properties are obtained using overall fits to all kaon data. New dedicated measurements would be welcome. 2) Perhaps the up-down quark mass difference is a factor of 2 or more larger than assumed in eq.~(\ref{eqtwentythree}). That would favor the smaller value of $|V_{us}| \lsim 0.220$ extracted from neutral $K_{e3}$ decays, a value that unfortunately does not seem to respect CKM unitarity.\cite{marcthree,quark,marcfour} 3) It is possible that the form factors $f^{K^0}_+$ and $f^{K^+}_+$ used in the extraction\cite{leut} of $|V_{us}|$ differ by more than the $m_d-m_u$ correction effect in eq.~(\ref{eqtwentythree}). 

How will the $K_{e3}$ puzzle be resolved? We will have to wait and see.
{\bf Note Added}: After the symposium, a more detailed study of radiative and chiral corrections to $K_{e3}$ decays appeared.\cite{cirig} It found an increase of about $+3\%$ for both decay modes, reducing $|V_{us}|$ by about 1.5\% for each. However, there is little change in eq.~(\ref{eqtwentyseven}) thus suggesting that the 9.5\% in eq.~(\ref{eqtwentytwo}) difference is of experimental origin and perhaps more likely in the neutral kaon system.

\section{Concluding Remarks}

I have tried to describe how lessons learned from our study of the muon lifetime in muonium can by analogy teach us something interesting about other bound state decay effects. Naive phase space decay rate reduction was shown to be canceled by final state interactions. Other effects such as $\mu^+e^-$ capture and quantum state reductions, although tiny for muonium, can have  significant analogous implications for muonic atoms and $b$-hadrons.

The examples discussed have interesting puzzles associated with them. The induced pseudoscalar coupling, $g_p$, in muon capture on hydrogen is too large. The $\Lambda_b$ lifetime is too short. The charged and neutral $K_{e3}$ decay rates appear to be in disagreement. Which one, if either,  gives the right value of $|V_{us}|$? Those types of puzzles are very healthy for physics. They stimulate new experiments and new ideas. Their resolution can lead to new discoveries and scientific or technological advances. It is the excitement of those discoveries and satisfaction of our intellectual appetites they provide that draws us to scientific research.  

On a personal note, I would like to end by expressing my gratitude for the work and discoveries of Vernon Hughes. His prized discovery, muonium, provided me with insights into bound state physics. He introduced me to the muon $g-2$. I had the pleasure to work on some theory related to his famous polarized $eD$ experiment at SLAC and a number of his other adventurous discoveries. He was the champion of muon physics. Hopefully, those of us who were inspired by his devotion to science will continue his tradition of excellence.

\end{document}